\documentclass[12pt]{iopart}

\usepackage{iopams}
\usepackage{graphicx}  
\begin{document}

\title{Efficient modeling of laser plasma interactions in high energy density scenarios}

\author{F Fiuza$^1$, M Marti$^1$, RA Fonseca$^{1,2}$, LO Silva$^1$, J Tonge$^3$, J. May$^3$, WB Mori$^3$}

\address{$^1$ GoLP/Instituto de Plasmas e Fus\~ao Nuclear - Laborat\'orio Associado, Instituto Superior T\'ecnico, Lisboa, Portugal}
\ead{frederico.fiuza@ist.utl.pt}

\address{$^2$ DCTI/ISCTE - Lisbon University Institute, Lisboa, Portugal}

\address{$^3$ Department of Physics \& Astronomy, University of California, Los Angeles, California 90095, USA}

\begin{abstract}
We describe how a new framework for coupling a full-PIC algorithm with a reduced PIC algorithm has been implemented into the code OSIRIS. We show that OSIRIS with this new hybrid-PIC algorithm can efficiently and accurately model high energy density scenarios such as ion acceleration in laser-solid interactions and fast ignition of fusion targets. We model for the first time the full density range of a fast ignition target in a fully self-consistent hybrid-PIC simulation, illustrating the possibility of stopping the laser generated electron flux at the core region with relatively high efficiencies. Computational speedups greater than 1000 times are demonstrated, opening the way for full-scale multi-dimensional modeling of high energy density scenarios and for the guiding of future experiments.
\end{abstract}

\maketitle

\section{\label{sec:intro} Introduction}

Laser-plasma interactions at ultrahigh intensities can lead to a wide range of nonlinear phenomena of interest for many applications such as fast ignition of inertial confinement fusion (ICF) targets \cite{bib:tabak, bib:atzeni, bib:kodama1, bib:kodama2}, plasma-based accelerators (of both leptons and hadrons) \cite{bib:dawson, bib:hegelich, bib:schwoerer, bib:silva1, bib:malka}, or laboratory astrophysics \cite{bib:bulanov}. Novel laser systems with unprecedented energies such as the National Ignition Facility \cite{bib:nif} and, in the near future, the HiPER project \cite{bib:hiper}, are coming online. In addition, short pulse lasers are now obtaining unprecedented peak intensities. These new high energy and high intensity lasers are now reaching the extreme conditions required for many of these applications. Numerical simulations also play a crucial role in understanding the physics and in optimizing the physical parameters and design for near and far term experiments. However, in several of these applications the diverse range of spatial and temporal scales involved make these numerical experiments extremely demanding in terms of computational resources. In some applications multi-dimensional full-scale simulations are not even yet possible to accomplish. As a consequence, many studies separate the simulations into fully electromagnetic (EM) kinetic studies of the microphysics at lower densities \cite{bib:sentoku1, bib:ren, bib:kemp, bib:tonge} and fluid/hybrid studies at higher densities \cite{bib:davies, bib:gremillet, bib:honrubia, bib:solodov}. The self-consistent coupling of the different scales has remained the holy-grail of fast and shock-ignition research.

Recently, a novel framework to coupling the disparate temporal, spatial, and density scales into an integrated simulation framework was proposed \cite{bib:cohen}. At low-density and high-temperature regions, where kinetic effects dominate, the full set of Maxwell's equations is solved as in standard PIC codes. At high-density low-temperature regions, where collisional effects dominate, thus damping EM and plasma waves, a reduced set of Maxwell's equations is solved, using an Ohm's law to calculate the resistive electric field, and the particles are pushed with the inclusion of a binary collision operator.

We have implemented this new framework into OSIRIS 2.0 \cite{bib:osiris}, a state-of-the-art PIC code, and we will discuss some of the implementation/numerical issues associated with the use of this numerical tool for the modeling of relevant scales and demonstrate its potential for full-scale modeling of the extreme scenarios associated with the laser-plasma interactions at ultrahigh intensities and densities. In particular, the possibility of performing for the first time full density scale modeling of fast ignition will be demonstrated. Our results open the way for full-scale multi-dimensional modeling of high energy density scenarios which can provide an integrated physical picture and clear directions for future experiments.

This paper is organized as follows. In Section \ref{sec:algorithm} we introduce the new algorithm, describe how it was implemented into OSIRIS 2.0, discuss some of the implementation details, and illustrate its accuracy and limitations. We will then apply the new algorithm to study the interaction of intense laser pulses with overdense targets in Section \ref{sec:sims}, concentrating on ion acceleration in laser-solid interactions and fast ignition. We thereby illustrate the potential of the new numerical scheme to do full-scale modeling of these scenarios. Finally, in Section \ref{sec:conclusions} we present our conclusions.

\section{\label{sec:algorithm} Algorithm}

Standard PIC codes \cite{bib:dawson2,bib:pic}, by solving the full set of Maxwell's equations, allow for an accurate and fully self-consistent modeling of laser-plasma interactions, providing a unique tool for the understanding of laser absorption and electron acceleration. By using an additional Coulomb collision module, capable of accurately describing Coulomb collisions between the different charged particles, PIC codes are also capable of accurately describing the electron transport from the critical density region up to solid densities ($\sim 10^{24}$ cm$^{-3}$) or even the ultrahigh densities of the core of a fast ignition fusion target ($\sim 10^{26}$ cm$^{-3}$). However, the need to resolve the very short characteristic temporal and spatial scales of the plasma at such high densities to avoid numerical instabilities presents a grand challenge. As a result full-PIC, full-density scale simulations of these regimes are not yet possible to accomplish. On the other hand, fluid/hybrid codes  \cite{bib:davies, bib:gremillet, bib:honrubia} use reduced models to describe the physics of electron transport and energy deposition at high densities and low temperatures, where collisional effects dominate. These reduced models allow for a considerable speedup of the simulations, by using larger spatial grids and time steps, but fail to model the laser-plasma interaction and the dominant kinetic effects associated with the low-density high-temperature region where electrons are accelerated, thus not allowing for a complete/self-consistent modeling of these scenarios. Most importantly they do not permit the self-consistent coupling between the two regions. The coupling of these two algorithms into a single framework proposed by Cohen et al \cite{bib:cohen} aims to solve this issue. In the low-density high-temperature region of the plasma, where kinetic effects dominate, the standard PIC algorithm is used, solving the full set of Maxwell's equations. In the high-density low-temperature region of the plasma, where collisions become dominant, and where EM and plasma waves are strongly damped, a reduced set of Maxwell's equations is solved using an inertialess electron equation of motion (or Ohm's law) to calculate the E-field and the standard Faraday's law to advance the B-field,
\begin{equation}
\mathbf{E} = \mathbf{\eta} \cdot \mathbf{J}_b - \mathbf{v}_e \times \mathbf{B}/c - (e n_e)^{-1} \nabla p_e + (e n_e)^{-1} \left(\frac{d}{dt} \right)_{coll, e-f} \mathbf{P}_e
\label{eq:Efield}
\end{equation}
\begin{equation} 
\frac{\partial \mathbf{B} }{\partial t} = - c \nabla \times \mathbf{E}
\label{eq:Bfield}
\end{equation}
where $\eta = m_e \nu_{ei}/n_e e^2$ is the classical resistivity \cite{bib:braginskii}, $\nu_{ei} \simeq 2\sqrt{2 \pi} e^4 Z_i^2 n_i/(3\sqrt{m_e} T_e^{3/2})$ ln$\Lambda$ is the classical electron-ion collision frequency, $Z_i$ is the ionization state, ln$\Lambda$ is the Coulomb logarithm, $T_e$ and $v_e$ are the background electron temperature and velocity, $J_b = J_e + J_i$ is the background current, which is the sum of the background electron current $J_e$ and ion current $J_i$, $e$ is the elementary charge, $n_{e,i}$ is the background electron or ion density, $p_e$ is the electron pressure, and the last term of Eq. \ref{eq:Efield} represents the momentum exchange due to collisions of fast electrons with background electrons. The different particle quantities can be calculated from the different moments of the background electrons.

We have implemented this new framework into OSIRIS 2.0 \cite{bib:osiris}, a fully relativistic, electromagnetic, and massively parallel, PIC code. In order to distinguish between background and fast electrons, we have followed the prescription of ref. \cite{bib:cohen}, considering an electron to be fast when its velocity exceeds a given multiple of the thermal velocity $\alpha v_{th}$; typically we use $\alpha = 5$ (we have verified that small variations on the exact value of the transition lead to very similar results). Using this prescription for defining fast electrons, the currents of the ions and of the fast electron, $J_i$ and $J_f$, can be deposited in the standard way and Amp\`ere's law, with the displacement current neglected, is used to calculate the background electron current as follows \cite{bib:cohen}
\begin{equation}
\mathbf{J}_e = (c/4\pi) \nabla \times \mathbf{B} - \mathbf{J}_i - \mathbf{J}_f .
\label{eq:ampere}
\end{equation}
By having both full-PIC and hybrid-PIC algorithms in the same framework, \emph{all} particles can be pushed in the standard way throughout the simulation, providing a natural distributed source for kinetic return currents. The use of a good collision operator capable of accurately describing the Coulomb collisions between plasma particles is crucial in order to guarantee the consistency between the microscopic description of the system (through inter particle collisions) and the macroscopic description (through the resistivity in Ohm's law). In our case, we use a relativistic corrected Monte Carlo binary Coulomb collision operator \cite{bib:peano} which generalizes previous works on binary Coulomb collision operators for PIC codes \cite{bib:abe,bib:nambu,bib:sentoku2} for arbitrary relativistic velocities, to ensure an accurate description of collisional effects and a smooth transition between full-PIC and hybrid-PIC regions.

\begin{figure}
\begin{center}
\includegraphics[width=0.9\textwidth]{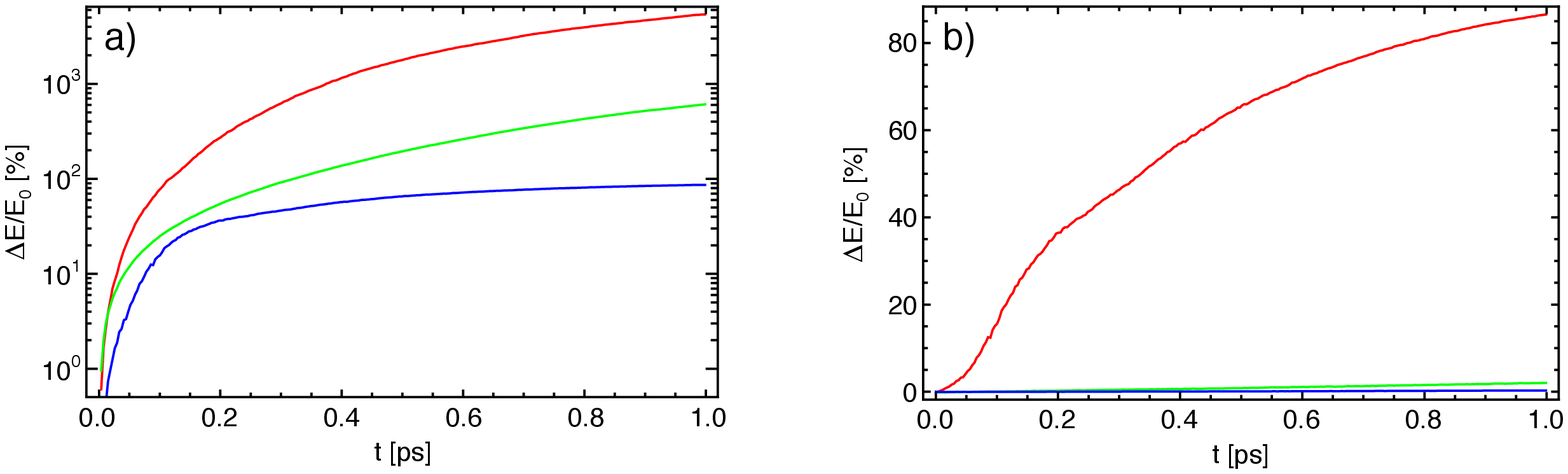}
\caption{\label{fig:numerics}(color online) Energy conservation in PIC codes for typical high energy density scenarios as a function of the particle interpolation scheme. The initial plasma density is $100 ~n_c$ and the initial temperature is 1 keV. a) After 1 ps, for linear interpolation (and smoothing plus compensator), numerical heating leads to a variation of $5400 \%$ of the energy in the simulation box with respect to the initial energy $E_0$ for $\Delta = 1.5 ~c/\omega_p$ and 16 ppc (red), $600 \%$ with $\Delta = 0.5 ~c/\omega_p$ and 16 ppc (green), and 87 \% with $\Delta = 1.5 ~c/\omega_p$ and 64 ppc (blue). b) Numerical heating can be dramatically improved using high-order splines. The increase of the energy in the simulation box using $\Delta = 1.5 ~c/\omega_p$ and 64 ppc is of $87 \%$ with linear (red), $2 \%$ with quadratic (green), and 0.3 \% with cubic interpolation (blue).}
\end{center}
\end{figure}

It is important to note that in order to have a smooth transition between the full-PIC and hybrid-PIC regions it is not enough to guarantee that the physical description is equivalent, and care must be given to ensure that the numerics are handled consistently on both sides of this transition. A critical issue is numerical grid heating from under-resolving the plasma Debye length, $\lambda_D$, which can occur in the full-PIC algorithm and does not occur in the hybrid-PIC one. In order to ensure a smooth transition the level of numerical heating must remain small in the full-PIC region, otherwise a strong temperature/resistivity gradient will occur at the transition region leading to different descriptions and inconsistent results. This issue can be overcome by using high-order splines for the particles and current smoothing, both of which are implemented in OSIRIS 2.0 \cite{bib:fonseca}. The effect of the different numerical parameters, as the number of particles per cell (PPC), the cell size ($\Delta$), and the particle shape, on the numerical heating/energy conservation in 2D PIC simulations for typical scenarios is illustrated in Fig. \ref{fig:numerics}. We have used current smoothing in all simulations. The cell size to a Debye length varies from 11-34 for the cases shown. We can observe that when using linear interpolation for the particles, even when the plasma skin depth, $c/\omega_p$, is resolved, and a large number of PPC is used, numerical heating takes place, largely increasing the energy of the system and potentially changing the physics in these high energy density scenarios (Fig. \ref{fig:numerics} a). The use of high-order splines can efficiently control/suppress numerical heating, even for $\Delta \gg \lambda_D$ (Fig. \ref{fig:numerics} b). Another important numerical effect that can degrade the transition between both algorithms has to do with the anomalous stopping of fast particles in plasmas when the number of macroparticles per skin depth cubed is small. For a small number of PPC and/or large cell size, the weight of each macroparticle becomes so large, leading to the generation of enhanced wakefields and therefore to enhanced energy exchange between fast particles and the plasma background that is artificially too large. Since this can only occur in the full-PIC algorithm (the hybrid algorithm does not support plasma waves) it can once more lead to different background temperatures on both sides of the transition. This effect will be described in more detail elsewhere \cite{bib:tonge2}. In our simulations we choose the number of PPC and cell size that allow for a good transition between the different algorithms. We use 1000 PPC in 1D and 64 PPC in 2D, with cell sizes of 0.5-1.5 $c/\omega_p$.

Since numerical constrains are significantly relaxed in the hybrid-PIC algorithm, the limiting cell size and time step is typically defined by the transition region between full-PIC and hybrid-PIC regions, where the local plasma skin depth and plasma frequency should be resolved. The transition between the full-PIC and hybrid-PIC algorithms is chosen according to the local density/collisionality of the plasma. The hybrid region map can be calculated each time step, dynamically following the evolution of the plasma density/temperature. For typical high-density low-temperature plasmas, the transition can be placed around 100 $n_c$, where $n_c$ ($\sim10^{21}$ cm$^{-3}$) is the critical density of the plasma for 1 $\mu$m light, leading to significant computational speedups, in comparison with full-PIC modeling, on the order of $(n_{max}/n_{hyb})^{(d+1)/2}$, where $n_{max}$ is the maximum density of the system, $n_{hyb}$ is the plasma density at the transition region, and $d$ is the dimensionality of the system. For simulations where significant regions at the maximum density exist the speedup is even larger due to the need of using larger number of PPC and finest resolutions in order to avoid the already mentioned numerical issues associated with the anomalous stopping of macroparticles in the full-PIC algorithm.

\begin{figure}
\begin{center}
\includegraphics[width=0.47\textwidth]{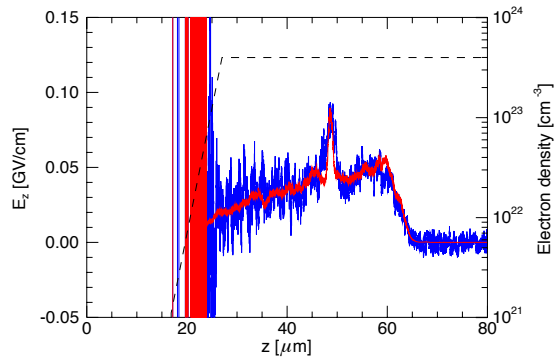}
\caption{\label{fig:hybrid}(color online) Comparison between the longitudinal E-field in full-PIC (blue) and hybrid-PIC (red) simulations performed with OSIRIS. A high-intensity laser pulse hits a cold (100 eV) plasma, with a density gradient from $10^{21}$ to $3.6 \times 10^{23}$ cm$^{-3}$ (dashed line), accelerating electrons forward and leading to the generation of a strong resistive E-field. The transition between full-PIC and hybrid-PIC in placed at $z = 25 ~\mu m$.}
\end{center}
\end{figure}

To illustrate the potential of this scheme, Fig. \ref{fig:hybrid} shows the resistive E-field for both a simulation with h-OSIRIS (OSIRIS hybrid-PIC simulations) and full-PIC OSIRIS one-dimensional (1D) simulations. In these simulations, a short laser pulse with a duration of 1 ps, a wavelength of $1 \mu$m, and an intensity of $5 \times 10^{19}$ Wcm$^{-2}$ (corresponding to a normalized vector potential $a_0 = 6$) interacts with a plasma with density of $360 ~n_c$ and an exponential density ramp from $n_c$ to $360 ~n_c$ with a scale length of $1.7 \mu$m. The transition between both algorithms is placed at 100 $n_c$ ($x_1 = 25 ~\mu$m). We can observe that the resistive field obtained in the hybrid region is in very good agreement with the E-field obtained from the full-PIC simulation, demonstrating the existence of a smooth transition between full-PIC and hybrid-PIC algorithms.

\section{\label{sec:sims} Simulations}

In order to demonstrate the potential of the new scheme for the modeling of scenarios where full-PIC or hybrid simulations isolated cannot give a self-consistent answer, we have applied this algorithm to the study of ion acceleration from laser-solid interactions and fast ignition of fusion targets.

\subsection{\label{subsec:ia} Ion acceleration in laser-solid interactions}

The interaction of a short and intense laser pulse with a solid-density target can lead to the generation of very energetic ions in short distances due to the strong sheath E-fields sustained by the plasma. As the laser pulse hits the plasma, it generates fast electrons that will propagate through the target and generate a strong quasi-static E-field ($\gtrsim 10$ GV/cm) at the back of the target due to the charge imbalance as electrons leave into vacuum. Modeling the interaction of the laser pulse with a solid density target becomes very demanding due to the need to model targets with thicknesses that range from 10s to 100s of $\mu$ms, to model acceleration times of several ps, and to resolve the plasma spatial and temporal scales at this high density. The typical solid targets used for ion acceleration have very low temperatures (10-100 eV) and therefore the transport of electrons is highly collisional. These conditions are therefore ideal for the application of the hybrid-PIC scheme. At the low-density region, where the laser pulse interacts with the plasma, the full-PIC algorithm is used, whereas at the high-density region the hybrid model is employed in order to model the highly collisional transport. At the back of the target, where the plasma density becomes low again, the full-PIC is used again to model the plasma expansion and ion acceleration in vacuum. 3D simulations of isolated targets are required in order to model ion acceleration in realistic scenarios that can be directly compared to experiments. Full-PIC simulations of these realistic conditions are not yet possible. Therefore, in order to compare our hybrid-PIC results with full-PIC simulations we resort to 1D and 2D simulations of ion acceleration.

\begin{figure}
\begin{center}
\includegraphics[width=0.9 \textwidth]{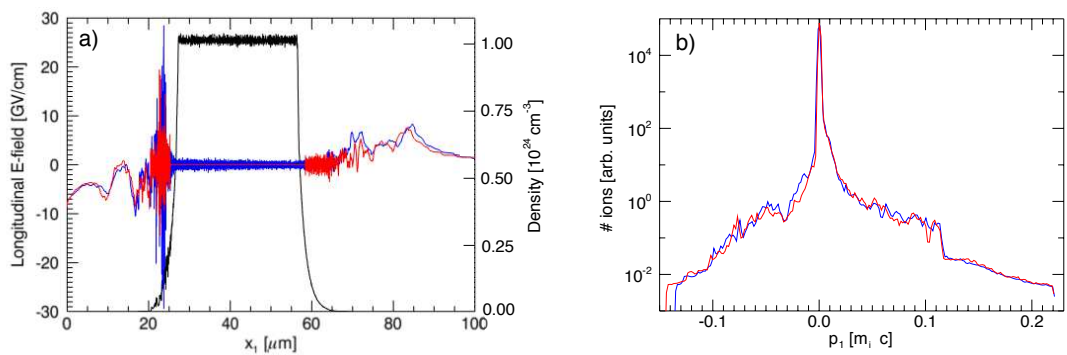}
\caption{\label{fig:comp_1d}(color online) Comparison of the (a) longitudinal E-field and (b) integrated ion spectrum in full-PIC and hybrid-PIC simulations after 0.85 ps. The hybrid-PIC curves (red) are in very good agreement with full-PIC results (blue). The plasma density profile corresponds to the black curve.}
\end{center}
\end{figure}

\begin{figure}
\begin{center}
\includegraphics[width=0.9 \textwidth]{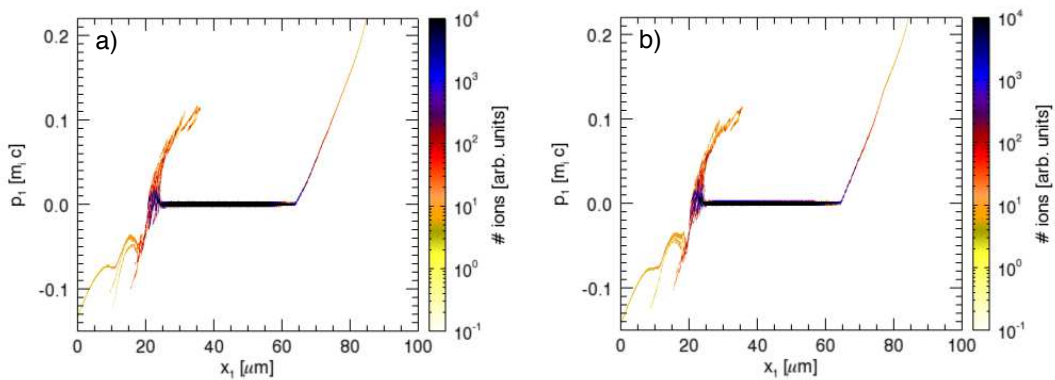}
\caption{\label{fig:p1x1_1d}(color online) Snapshot of the ion longitudinal momentum versus longitudinal position in (a) full-PIC and (b) hybrid-PIC simulations after 0.85 ps.}
\end{center}
\end{figure}

\begin{figure}
\begin{center}
\includegraphics[width=0.9 \textwidth]{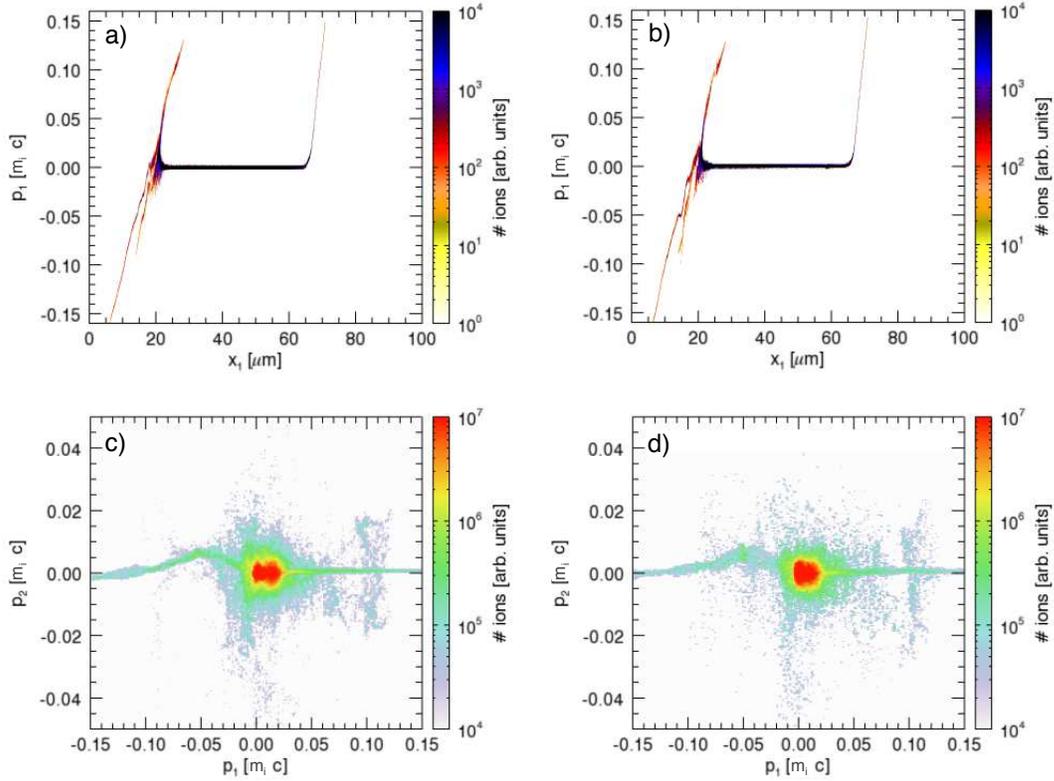}
\caption{\label{fig:ps_2d}(color online) Snapshots of the ion phase-space from 2D full-PIC and hybrid-PIC simulations after 0.5 ps. Top panels: longitudinal momentum versus longitudinal position in (a) full-PIC and (b) hybrid-PIC simulations. Bottom panels: transverse momentum versus longitudinal momentum in (c) full-PIC and (d) hybrid-PIC simulations.}
\end{center}
\end{figure}

In our simulations a short laser pulse, with a gaussian temporal profile and a duration of 350 fs in 1D and 200 fs in 2D, and with a peak intensity of $5 \times 10^{19}$ Wcm$^{-2}$, interacts with a solid target with a $30 ~\mu$m thickness, and with a density and temperature of $1000 ~n_c$ and 100 eV in 1D and $500 ~n_c$ and 30 eV in 2D. A plasma ramp is placed on each side of the target in a density gradient with a scale length of $1.7 ~\mu$m, to take into account the plasma expansion from a prepulse. The longitudinal E-field along the simulation box and the integrated ion spectrum are in very good agreement between OSIRIS and h-OSIRIS simulations (Fig. \ref{fig:comp_1d}). Ions reach a maximum energy of 22.5 MeV after 0.85 ps. A more detailed analysis of the ion phase-space for both OSIRIS and h-OSIRIS simulations reveals that all the features of ion acceleration are recovered by the hybrid-PIC simulations both in 1D (Fig. \ref{fig:p1x1_1d}) and in 2D (Fig. \ref{fig:ps_2d}). The hybrid-PIC simulations were 90 times faster than the full-PIC simulations in 1D and 300 times faster in 2D and still able to accurately describe all the relevant physics, demonstrating the potential of using the new algorithm for efficient full-scale modeling of ion acceleration in solid targets.

\subsection{\label{subsec:fi} Fast ignition}

\begin{figure}
\begin{center}
\includegraphics[width=0.47\textwidth]{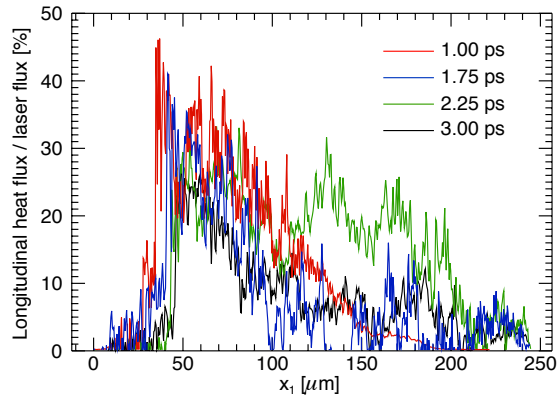}
\caption{\label{fig:fi_full}(color online) Evolution of the longitudinal electron heat flux in a full-PIC simulation of a fast ignition-like target with maximum density of 500 $n_c$. The initial heat flux ($\sim 30 \%$ of the laser energy flux) is transported along the target, not being absorbed due to the low collisionality, and refluxing around the target, as illustrated by the strong variations in time.}
\end{center}
\end{figure}

Fast ignition of fusion targets is another important application where full-scale modeling is crucial. Developing a complete understanding of the underlying physics of fast ignition from laser absorption to the energy deposition in the core region represents a necessary step towards the realization of such scheme for high-gain ICF. Due to the high degree of complexity of the physics involved and the difficulty in performing ignition relevant experiments with current facilities, this understanding relies heavily in numerical simulations. Realistic self-consistent full-PIC simulations of fast ignition require the modeling of compressed targets with radii of 100s $\mu$m, and modeling ignition lasers with duration $> 10$ ps, while resolving the plasma skin depth and time scale at the core, which has a density of $\sim 10^{26} $ cm$^{-3}$, corresponding to $c/\omega_p \sim 5$ \AA  ~and $\omega_p^{-1} \sim 2$ as. The tremendous difference in the scales involved makes this modeling impractical and therefore most PIC simulations of fast ignition are limited to densities $<100 ~n_c$, plasma lengths $<50 ~\mu$m, and laser durations/propagation times $<1$ ps. Apart from the numerical issues that were already discussed, in these reduced scale simulations the plasma length is relatively small and the density is not high enough to stop the majority of fast electrons, leading to a strong fast electron current crossing the back wall of the simulation box, where particles will be absorbed or thermally re-emitted leading to a jump in current and to the build up of a strong E-field. This E-field will reflect and reflux particles, leading to a hot return current, increasing the plasma pressure and changing laser absorption at the vacuum-plasma interface, electron acceleration and transport. Even in the case where the plasma is isolated from the walls of the simulation box, fast particles will still reflux in the target due to the strong space-charge field that builds up at the back of the target. This is illustrated in Fig. \ref{fig:fi_full}, where the longitudinal electron heat flux is shown for different simulation times, up to 3 ps. In this 1D full-PIC simulation, a laser with a peak intensity of $5 \times 10^{19}$ Wcm$^{-2}$ is used, having a rise time of $500$ fs and then a constant profile at the maximum intensity. The plasma has an exponential density ramp on both sides from $n_c$ up to $500 ~n_c$, with a scale length of $6.8 ~\mu$m, and has a flat density region at $500 ~n_c$ of $112 ~\mu$m. At this moderate density, the collisionality is not high enough to stop most fast electrons, and therefore they cross the target refluxing at the back, leading to a hot return current and strongly modifying the heat flux in time. In order to avoid this artificially strong refluxing it is crucial to model fast ignition targets with a realistic density profile, where particles are self-consistently generated and absorbed at the core, and where no artificial return current is present which could modify the interaction.

In a fast ignition compressed target, the different physical regimes associated with the varying densities and temperatures occur in different regions. In the coronal region, with lower densities ($\sim n_c$) and higher temperatures ($\sim$ keV), where the ignition laser interacts with the plasma, kinetic effects dominate, strong EM and plasma waves can be present and full-PIC modeling is ideal. Close to the core region, with very high densities ($\sim 10^{26} $ cm$^{-3}$) and low temperatures ($\sim 10-100$ eV), collisional effects dominate, strongly damping EM and plasma waves, and allowing the use of reduced models. This separation is ideal for the use of the new hybrid-PIC scheme, where the full-PIC algorithm is used to solve the interaction of the ignition laser with the coronal plasma and electron acceleration and transport up to moderate densities, and the hybrid-PIC algorithm is used to solve the transport and energy deposition of fast electrons from moderate densities to the core.

In order to demonstrate the possibility of studying the self-consistent interaction of the ignition pulse with a close to realistic fast ignition fusion target, we have performed h-OSIRIS simulations modeling for the first time the full density range of a fast ignition target. Using the same setup of the simulations presented in Fig. \ref{fig:fi_full}, we have extended the plasma density up to $10^{5} n_c$ ($\sim10^{26}$ cm$^{-3}$), using the same density gradient length scale and having a core region of $40 ~\mu$m. The self-consistently generated fast electron population is now stopped at the very dense core due to collisions, causing it to heat up. The temporal evolution of the longitudinal heat flux denotes a relatively constant flux carrying $\sim 30 \%$ of the laser energy flux. After 3 ps, $\sim 17 \%$ of the laser energy has been deposited into ions; however, the majority of this energy was deposited before the core, with only $5 \%$ of the laser energy ($17 \%$ of the electron heat flux) being deposited into the ions at the core. This number can certainly be improved with a better, and more realistic, target profile or with the inclusion of a high-Z material cone, where the laser is absorbed closer to the core and the density gradient is steeper, shortening the distance between $10^{25} - 10^{26}$ cm$^{-3}$ densities, where most of the energy is deposited into ions ($12 \%$ of the laser energy flux). We also note that on one hand, significantly higher absorption efficiencies are expected in multi-dimensions, since typically absorption efficiencies $> 60\%$ of the laser energy flux are observed in 2D \cite{bib:kemp,bib:tonge}. On the other hand, in multi-dimensions fast electron divergence will lead to a lower efficiency at the core, depending on the divergence angle and distance to the core. The importance of these multi-dimensional effects in realistic fast ignition targets will be addressed in more detail elsewhere \cite{bib:fiuza}. The full density range hybrid-PIC simulation performed with h-OSIRIS was $> 1000$ times faster than a full-PIC simulation of the same setup would have been.

\section{\label{sec:conclusions} Conclusions}

A new framework that couples a full-PIC code with a hybrid-PIC algorithm has been implemented into OSIRIS 2.0, allowing for the modeling of laser plasma-interactions in scenarios where high densities and large density gradients are present. We discussed the implementation and the limitations associated with this new scheme, with an emphasis on the numerical consistency between the two algorithms, illustrating that a smooth/consistent transition can be achieved. Hybrid-PIC simulations of ion acceleration in laser-solid interactions have been performed and compared with full-PIC simulations both in 1D and 2D, showing that the hybrid-PIC algorithm can be highly efficient and accurate. We have also discussed some of the issues associated with the full-PIC modeling of large density gradients at reduced scales in the context of fast ignition. Without the inclusion of a dense core, a strong refluxing can occur due to the inefficiency of a moderate density plasma in stopping fast electrons \cite{bib:tonge}. This can lead to artificially hot return currents, which can change the interaction conditions. While refluxing can be reduced through the use of an absorbing core \cite{bib:tonge}, the incorrect heat capacity of the core can lead to large return currents. These results confirm the need to simulate the full density range of a fast ignition target in order to accurately characterize ignition conditions.

Using the hybrid-PIC framework, we have modeled for the first time the full density range of associated with a fast ignition target, demonstrating the possibility of studying in a self-consistent manner the different physics involved, from laser absorption and electron acceleration, to electron transport and energy deposition in the core. Our preliminary 1D results show that the laser generated electron flux is strongly stopped at the core due to collisions, leading to relatively high conversion efficiencies from fast electrons to core ions (~$17 \%$), which is crucial in order to achieve ignition conditions \cite{bib:atzeni}. The appropriate shaping of the fast ignition target or the inclusion of a high-Z material cone might lead to even higher efficiencies, since a substantial fraction of the energy carried by fast electrons is observed to be deposited before the core. Computational speedups greater than 1000 times have been achieved, opening the way for full-scale multi-dimensional modeling of high energy density scenarios and for the guiding of future experiments.

\section*{Acknowledgments}
The authors acknowledge useful discussions with B. I. Cohen and A. J. Kemp. Work partially supported by FCT (Portugal) through the grants PTDC/FIS/66823/2006 and SFRH/BD/38952/2007, by the European Community (HiPER project EC FP7 no. 211737), and by the NSF under Grant Nos. PHY-0904039 and PHY-0936266, and by the DOE under Contract Nos. DE-FG03-09NA22569 and under the Fusion Science Center for Matter Under Extreme Conditions. The simulations were performed at the Intrepid supercomputer (ANL), the Jugene supercomputer (Germany), the Hoffman cluster (UCLA), and the IST cluster (Lisbon, Portugal).

\newpage

\begin{figure}
\begin{center}
\includegraphics[width=0.47\textwidth]{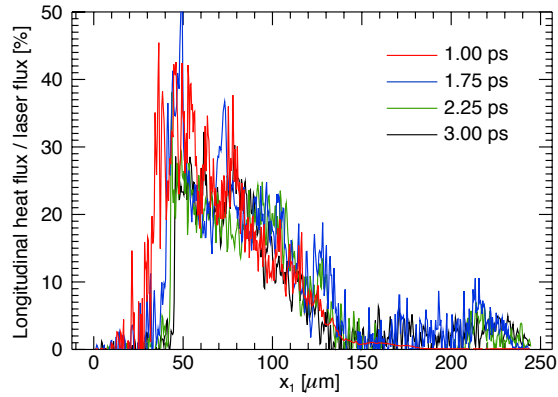}
\caption{\label{fig:fi_hybrid}(color online) Evolution of the longitudinal electron heat flux in a hybrid-PIC simulation of the full-density range of a fast ignition target. The initial heat flux ($\sim 30 \%$ of the laser energy flux) is strongly stopped due to collisions as electrons reach the high density core region. The heat flux is approximately constant in time.}
\end{center}
\end{figure}

\begin{figure}
\begin{center}
\includegraphics[width=0.47\textwidth]{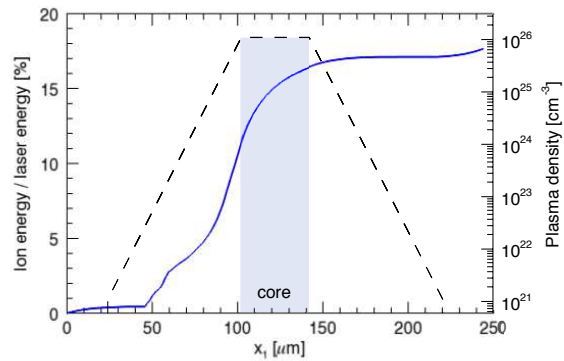}
\caption{\label{fig:fi_hybrid}(color online) Integrated ion energy (solid line) as a function of the target position. After 3 ps, $\sim 17 \%$ of the laser energy has been deposited into the plasma ions, 11 $\%$ before the core, 5 $\%$ at the core (light-blue region), and 1$\%$ after the core. The energy deposited into the ions at the core corresponds to $\sim 17 \%$ of the electron heat flux generated by the laser. The dashed line represents the initial plasma density profile. }
\end{center}
\end{figure}

\end{document}